\documentclass[lettersize,journal]{IEEEtran}
\usepackage{amsmath,amsfonts} 
\usepackage{amssymb} 
\usepackage{MnSymbol}
\usepackage{extarrows}
\usepackage{enumerate} 
\usepackage{bm} 
\usepackage{booktabs}
\usepackage{array}

\usepackage{graphicx} 
\usepackage{subfigure} 
\usepackage{multirow} 

\usepackage{cite}
\usepackage{algorithmic}
\usepackage{algorithm}
\usepackage{array}
\usepackage[caption=false,font=normalsize,labelfont=sf,textfont=sf]{subfig}
\usepackage{textcomp}
\usepackage{stfloats}
\usepackage{url}
\usepackage{verbatim}
\usepackage{xcolor}
\usepackage{graphicx}
\usepackage{epstopdf}
\newtheorem{theorem}{Theorem}[section]
\newtheorem{definition}{Definition}[section]

\newtheorem{lemma}{Lemma}[section]

\newtheorem{proposition}{Proposition}[section]
\newtheorem{corollary}{Corollary}[section]
\newtheorem{remark}{Remark}[section]
\newtheorem{example}{Example}[section]

\newcommand{\pf}{\textbf{Proof: }}
\newcommand{\e}{\hfill$\blacksquare$}

\allowdisplaybreaks[4] 

\date{}
\def\BibTeX{{\rm B\kern-.05em{\sc i\kern-.025em b}\kern-.08em
		T\kern-.1667em\lower.7ex\hbox{E}\kern-.125emX}}
\usepackage{balance}
\begin{document}

	\title{Partial Orders in Rate-Matched Polar Codes
	}

	\author{\fontsize{10.3pt}{\baselineskip}\selectfont Zhichao Liu$^{*\dagger}$, Liuquan Yao$^{*\dagger}$,  Yuan Li$^{\ddagger}$, Huazi Zhang$^\ddagger$,  Jun Wang$^\ddagger$, Guiying Yan$^{*\dagger}$ and Zhiming Ma$^{*\dagger}$\\
		$^*$University of Chinese Academy and Sciences, Beijing, China\\
		$^\dagger$Academy of Mathematics and Systems Science, CAS, Beijing, China\\ 
		$^\ddagger$Huawei Technologies Co. Ltd., China\\

		Email: \{liuzhichao20, yaoliuquan20\}@mails.ucas.ac.cn, \{liyuan299, zhanghuazi, justin.wangjun\}@huawei.com,\\
		yangy@amss.ac.cn, 	mazm@amt.ac.cn
		\thanks{This work was supported by the National Key R\&D Program of China (No. 2023YFA1009602).}
	}

	
	\maketitle
	
	\begin{abstract}
        In this paper, we establish the partial order (POs) for both the binary erasure channel (BEC) and the binary memoryless symmetric channel (BMSC) under any block rate-matched polar codes. Firstly, we define the POs in the sense of rate-matched polar codes as a sequential block version. Furthermore, we demonstrate the persistence of POs after block rate matching in the BEC. Finally, leveraging the existing POs in the BEC, we obtain more POs in the BMSC under block rate matching. Simulations show that the PW sequence constructed from $\beta$-expansion can be improved by the tool of POs. Actually, any fixed reliable sequence in the mother polar codes can be improved by POs for rate matching.
	\end{abstract}
	\section{Introduction}\label{section1}
	
	\IEEEPARstart{S}{ince} Ar{\i}kan's introduction of polar codes \cite{arikan2008}, polar codes have garnered significant attention and research interest, which are not only capacity achieving, but also have encoding and decoding algorithm with low complexity. In traditional polar codes, as the code length $N$ approaches infinity, the ratio of perfect channels converges to the channel capacity $I(W)$, while the ratio of purely noisy channels approaches $1-I(W)$. Thus, a crucial aspect in polar code research is the selection of the $K$ most reliable bit channels from $N$ synthetic channels to carry information bits. For traditional polar codes, there are many methods to select information bits, such as Gaussian approximation (GA) algorithm \cite{b11}, PW construction algorithm \cite{beta-expansion} and 5G sequence \cite{b21}.
	
	In engineering applications, the code length $N$ typically does not reach exponential powers. To address this, the issue of rate matching has been introduced in \cite{coding} and \cite{coding2}. Generally, a bit-puncturing strategy is employed at lower rates, while a bit-shortening strategy is employed at higher rates. Recently, it has been demonstrated that puncturing and shortening polar codes achieve capacity in \cite{ca_ac}. There have been many studies related to rate matching and refer to \cite{ca_ac}-\cite{8692789} for more details. Although we can use algorithms such as Gaussian approximation to reconstruct the bit reliability of the rate-matched polar codes, the corresponding complexity is still very high. While if we can obtain a family of POs, we can avoid some repeated reliable calculations under different SNR. Therefore, it remains highly significant to investigate the POs of rate-matched polar codes.
	
	Because the analysis of the Bhattacharyya parameters has recursive expressions for BEC, many researches study the POs in the BEC \cite{wang1} \cite{wang2} \cite{wang3}. Channel degradation has been proposed in \cite{channel1}, then in \cite{channel2} they verify the partial order between $10$ and $01$ under channel degradation by a sufficient condition. What's more, \cite{general} studies the generalized partial order of BEC and BMSC. However, the analysis after rate matching in the BEC and BMSC is complex because the reliability of positions changes without recursion. Despite its engineering significance, this issue has received little attention. As we know, we firstly propose the POs for both the BEC and the BMSC under sequential rate-matched polar codes.
	
    In \cite{newpo}, we explored new POs for traditional polar codes by analyzing the upper and lower bounds of the Bhattacharyya parameters in BMSC. We find more POs defines by Bhattacharyya parameter and error probability, which are richer than POs in the sense of channel degradation. Building upon this work, our objective is to find more POs in the BEC and BMSC under sequentially punctured polar codes.
	\subsection{Contributions}
	The contributions of our paper are summarized as follows:
	\begin{enumerate}
	\item We firstly propose a study of partial order under block rate matching, and define the partial order of BEC and BMSC under block rate matching. This has a certain enlightenment effect on the ranking of the reliability of polar codes after rate matching.\\
	\item We extend some inherited POs, and propose a sufficient condition for verifying the inherited POs of the BEC under block rate matching. Since the partial order of the mother polar code under the BEC is rich, this research provides a certain guarantee for the diversity of the partial order of the BEC under rate matching.\\
	\item We introduce novel POs for the BMSC under block rate matching. Our conclusion leverages the existing POs for BEC under block rate matching to derive new POs applicable to arbitrary BMSC also under the same block rate matching. The proof constitutes a central challenge of our work. In order to obtain an inequality of the Bhattacharyya parameters of the synthesized channel after rate matching, our approach constructs a convolution mapping and demonstrates that the geometric mean pair exhibits a superior polarization effect compared to the uneven pair of initial Bhattacharyya parameters.\\
	\end{enumerate}
	We have listed the main conclusions of our paper in Table \ref{table 1}.
	
	%
	%
		\begin{table}[t]
		\caption{Inherited BEC POs and POs for BMSC}
		\begin{center}
			\begin{tabular}{|c|c|}
				\hline
		Conditions & POs\\
		        \hline
				 $a \preceq_{P,m,BEC} b,|a|,|b|\ge m$ & $ac \preceq_{P,m,BEC} bc $\\
		        \hline
				 $a \preceq_{P,m,BEC} b, c\preceq_{BEC} d,|a|,|b|\ge m$ & $ac \preceq_{P,m,BEC} bd $\\
		        \hline
				 $n,t,\ell \ge 0$ & $p^n 0r^\ell 1 q^t \preceq_{P,m,BEC} p^n 1r^\ell 0 q^t $\\
		        \hline
				 $0^p 1^{m-p} \alpha \preceq_{BEC} 0^q 1^{m-q} \gamma$, $q\le p $&$0^p 1^{m-p} \alpha \preceq_{1,m,BEC} 0^q 1^{m-q} \gamma $\\
				\hline
					$\mid \gamma \mid =m-1$, $\gamma 0 \alpha  \preceq_{BEC}  \gamma 1 \beta $&$\gamma 0 \alpha  \preceq_{P,m,BEC}  \gamma 1 \beta$\\
				\hline
				$\gamma \preceq_{P,m,BEC} \alpha   $&$\gamma 1\preceq_{P,m,BMSC} 1\alpha$\\
				\hline
			\end{tabular}
			\label{table 1}
		\end{center}
	\end{table}
	
	\subsection{Organizations and Notations}
	This paper is organized as follows. We review the definition of polar codes and rate matching, and define the partial order under block rate matching in Section \ref{section2}. In Section \ref{section3}, we give some inherited POs in the BEC and a sufficient condition for block rate matching. In Section \ref{section4}, we give the details of the proof of the new POs in BMSC along with an illustrative example. Simulation results are presented in Section \ref{section5}. Finally, conclusions are drawn in Section \ref{section6}. 
	
	Greek letters denote the paths of polarization, where '$0$' represents up polarization and '$1$' represents down polarization. The modulo length $|\cdot|$ indicates the length of the path.
	
	For example, $\alpha=1100$ represents a path involving two down polarization transformations followed by two up polarization transformations. We use $W^\alpha$  to denote the synthesized channel generated by the polarization path $\alpha$ with the initial channel $W$. In this paper, we exclusively consider $W$ as either a BEC or BMSC.
	
	For consistent initial input variables, we define the up and down polarization functions as follows: $f_0(x):=1-(1-x)^2, f_1(x):=x^2$ with $x\in [0,1]$, and then for a path $\alpha$ with length $n$, we define
	\begin{equation}\label{bp}
	\begin{aligned}
	f_\alpha(x)&=f_{\alpha_1\alpha_2\cdots\alpha_n}(x)\\
	&:=f_{\alpha_n}\circ f_{\alpha_{n-1}}\circ\cdots\circ f_{\alpha_1}(x),\;\; x\in[0,1].
	\end{aligned}
	\end{equation}
	
	Generally, we define up and down polarization operations as '$\bar{*}$' and '$\underline{*}$' relatively for different input variables:
	
	\begin{equation}\label{punc_evo}
	\left\{
	\begin{aligned}
	& a\bar{*}b=a+b-ab,\\
	& a\underline{*}b=ab.\\
	\end{aligned}
	\right.
	\end{equation}
	
	For a vector $Z$ of length $N=2^n$, where $Z=(z_1,\cdots,z_n)$, we denote the polarized vector of $Z$ as $h(Z)$. For example, when $Z=(z_1,z_2,z_3,z_4)$, then
	
\begin{equation}
\begin{aligned}
h(Z)= & [(z_1 \bar{*} z_3)\bar{*}(z_2\bar{*}z_4),(z_1 \bar{*} z_3)\underline{*}(z_2\bar{*}z_4),\\
& (z_1 \underline{*} z_3)\bar{*}(z_2\underline{*}z_4),(z_1 \underline{*} z_3)\underline{*}(z_2\underline{*}z_4)]\\
\end{aligned}
\end{equation}	

Fig. 1 illustrates the polarized process of different input variables.

\begin{figure}[!t]
		\centering
		\includegraphics[width=7.5cm,height=4.0cm]{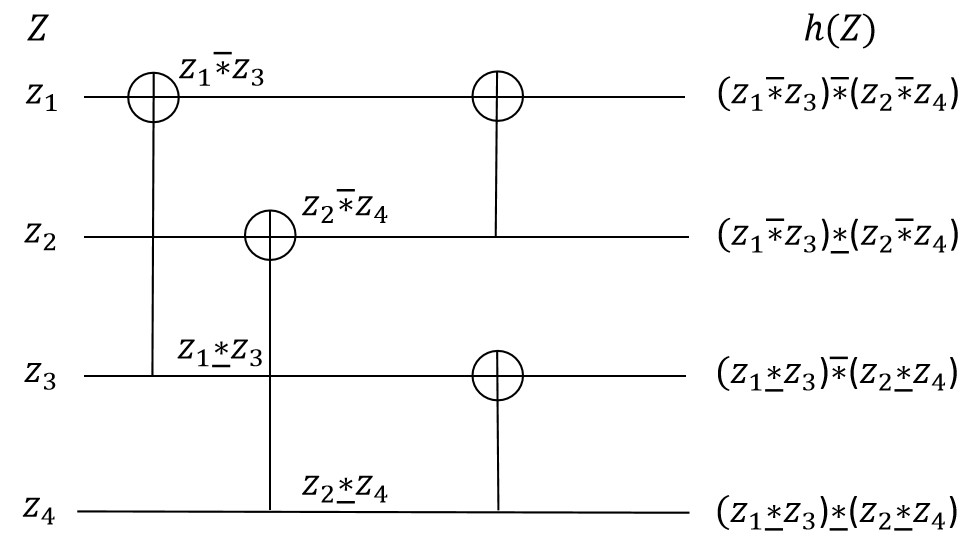}
		\caption{$h(Z)$ as the polarized vector of $Z$ for $N=4$. }
	\end{figure}

$h(Z_1)\le h(Z_2)$ indicates that every component of the vector satisfies the inequality.
	
    \section{Preliminaries}\label{section2}	
	\subsection{Polar Codes}
	Let $N=2^n$,$F_N=B_N F_2^{\otimes n}$, where $F_2=\left(\begin{aligned}
& 1& 0\\
& 1& 1\\
\end{aligned}\right)$ and $B_N$ is the bit-reversal permutation matrix. Then a $\mathcal{P}(N,K)$ polar code can be generated by choosing $K$ rows in $F_N$ as the information bit set $\mathcal{I}$, and other rows are denoted by $\mathcal{F}$. The transmitted code can be encoded by $\boldsymbol{x}=\boldsymbol{u} F_N$, where only $\boldsymbol{u}(\mathcal{I})$ can carry information while $\boldsymbol{u}(\mathcal{F})=0$. The codeword $\boldsymbol{x}$ is transmitted through a channel and successive cancellation (SC) decoder \cite{arikan2008} is a frequently used decoder algorithm with low complexity $O(NlogN)$. In order to improve the performance of decoder, a successive cancellation list (SCL) decoder is proposed in \cite{list}. And a SCL decoder with CRC is studied in \cite{crc} for detecting the error in the list.
	\subsection{Rate Matching}
	The code length of the polar code is an integer power of two: $N=2^n$, but the code length is usually required to be an arbitrary positive integer. Therefore, rate matching is a method used to modify the code length and adjust rate. The topics we consider for rate matching include block puncturing and sequential shortening.
	\begin{enumerate}
	\item $Puncturing:$ Puncturing makes $\bar{P}$ bits incapable to modify the code length, and the punctured code length is $N-\bar{P}$. After the decoder receives the received vector, it has no punctured bit information at all. Thus, the LLRs of the punctured bit is $0$ and the Bhattacharyya parameter is $1$. Block puncturing means sequentially puncturing out the encoded $1,2,\cdots, \bar{P}$ bits.\\
	\item $Shortening:$ Shortening makes $\bar{S}$ bits fixed, and when the decoder receives the received vector, it is fully aware of the information about the shortened bits. Therefore, the LLR of the shortened bits is $1$ and the Bhattacharyya parameter is 0. Block shortening means shortening the encoded $N-\bar{P}+1,\cdots,N $ bits from the last bit.\\
	\end{enumerate}
	\subsection{Definitions of PO under Block Rate Matching}
	In this subsection, we define the POs for the BEC and the BMSC under block rate matching. 

    Firstly, we provide an equivalent description of any block puncturing and shortening:
\begin{definition}
Given an odd integer $P$ and $ m\in \mathbb{N}$, s.t. $m>log_2 P$. For $\forall$ $N=2^n \ge 2^m$, we define $P/2^m$ puncturing where we puncture $\frac{P}{2^m} N$ bits in the positions $\{1,\cdots,\frac{P}{2^m} N\}$ from a $N$ length polar code. 
\end{definition}    
\begin{definition}
Given an odd integer $S$ and $ m\in \mathbb{N}$, s.t. $m>log_2 S$. For $\forall$ $N=2^n \ge 2^m$, we define $S/2^m$ shortening where we shorten $\frac{S}{2^m} N$ bits in the positions $\{N-\frac{S}{2^m} N+1,\cdots,N\}$ from a $N$ length polar code. 
\end{definition}

We provide an illustrative example explaining the similarity of different code length $N$ under $P/2^m$ puncturing.
    
\begin{example}  
In the case of $1/4$ puncturing, $N=4$ and $N=8$ result in the puncturing of $1$ and $2$ bits relatively. And we describe the relationship between them above.
    
    Let $Z=(1,\epsilon,\epsilon,\epsilon)$ and $Y=(1,1,\epsilon,\epsilon,\epsilon,\epsilon,\epsilon,\epsilon)$ are the initial Bhattacharyya parameter for $N=4$ and $N=8$ relatively, where $\epsilon$ is the erasure probability of channel.Then 
    
    \begin{equation}
    \begin{aligned}
    h(Z)& \overset{\triangle}{=}(h(Z)_1,h(Z)_2,h(Z)_3,h(Z)_4)\\
    & =(1,2\epsilon-\epsilon^2,\epsilon+\epsilon^2-\epsilon^3,\epsilon^3).\\
    \end{aligned}
    \end{equation}
    
    And $h(Y)$ can be derived from $h(Z)$:
    
    \begin{equation}
    \begin{aligned}
    h(Y)= & (2h(Z)_1-h^2(Z)_1,h^2(Z)_1,\\
    & 2h(Z)_2-h^2(Z)_2,h^2(Z)_2,\\
    & 2h(Z)_3-h^2(Z)_3,h^2(Z)_3,\\
    & 2h(Z)_4-h^2(Z)_4,h^2(Z)_4).\\
    \end{aligned}
    \end{equation}
\end{example}  
    Consequently, the evolution rules of the Bhattacharyya parameter in terms of the simplest fraction $P/2^m$ exhibit similarities. This is the reason why we define sequential puncturing in this way.

The definitions of the Bhattacharyya parameters of the synthesized channel after block rate matching are given for BEC and BMSC.
\begin{definition}
For an initial BEC $W$ with polarization path $\alpha$, define $Z_{P,m,\alpha}(x)$ be the polarized Bhattacharyya parameter under $P/2^m$ puncturing and $Z_{S,m,\alpha}(x)$ be the polarized Bhattacharyya parameter under $S/2^m$ shortening.
\end{definition}

\begin{definition}
For initial BMSC $W$, define $Z_{P,m}(W^\alpha)$ be the polarized Bhattacharyya parameter of channel $W^\alpha$ under $P/2^m$ puncturing and $Z_{S,m}(W^\alpha)$ be the polarized Bhattacharyya parameter of channel $W^\alpha$ under $S/2^m$ shortening.
\end{definition}	

	Then, we establish the POs based on the Bhattacharyya parameter.
    
    \begin{definition}
		We write $\alpha\preceq_{P,m,BEC} \gamma$ iff
		\begin{equation}
		Z_{P,m,\gamma}(x)\le Z_{P,m,\alpha }(x),\;\;\forall x\in [0,1].
		\end{equation}
		
        We write $\alpha\preceq_{P,m,BMSC} \gamma$, iff
        
        \begin{equation}
		Z_{P,m}(W^\gamma)\le Z_{P,m}(W^\alpha ),\;\;\forall x\in [0,1].
		\end{equation}
		The definitions of $\alpha\preceq_{S,m,BEC} \gamma$ and $\alpha\preceq_{S,m,BMSC} \gamma$ under shortening are the same as puncturing.
		
	\end{definition}
    
    For $\alpha=\alpha_1 \cdots \alpha_t$, $Z_{P,m,\alpha }(x)$ can be calculated by $t$ iterations of either $\bar{*}$ or $\underline{*}$ convolutions, as illustrated in Fig. 1. When the convolution layer $n$ exceeds $m$ which means $m<n\le t$, the evolution can be simplified to the traditional case instead of (\ref{punc_evo}):
    
    \begin{equation}
    \left\{
    \begin{aligned}
    & Z_n=2Z_{n-1}-Z_{n-1}^2,\ \alpha_n=0\\
    & Z_n=Z_{n-1}^2,\ \alpha_n=1\\
    \end{aligned}\right.
    \end{equation}
    
    where $Z_n$ denotes the Bhattacharyya parameter under $n$-th layer of $\alpha$. Specially, $Z_0$ denotes the initial Bhattacharyya parameter of $\alpha$ and $Z_t=Z_{P,m,\alpha }(x)$.
    
    \section{Inherited Partial orders in the BEC under block rate matching}\label{section3}
    We give the inherited POs under the BEC in two parts. In the first part, we get some POs similar to those under the mother polar code. And in the second part, utilizing a sufficient condition, we deduce two conclusions.
    \subsection{Some Conclusions Similar to the Old POs for the Mother Polar Codes}
Firstly, we give some recursive rules of the POs in the BEC under block rate matching:
    \begin{proposition}
    Consider the $P/2^m$ puncturing and $S/2^m$ shortening, and given strings $a$, $b$ satisfied $|a|\ge m$, $|b|\ge m$, for any strings $c$, $d$ ,
    
    (i) if $a\preceq_{P,m,BEC} b$, then $ac\preceq_{P,m,BEC} bc$;
    
    (ii) if $a\preceq_{P,m,BEC} b$ and $c\preceq_{BEC} d$, then $ac\preceq_{P,m,BEC} bd$.
    
    This is also true for $\preceq_{S,m,BEC}$.
    \end{proposition}
    \pf (i) is because $f_c(x)$ is monotonically increasing; use (i) and $Z_{P,m,b}(x)\in [0,1]$, then we get (ii).\e  
    
    Then we deduce the relationship between $10$ and $01$. It is then proved that under the BEC, inserting arbitrarily identical sequences before, after, and in the middle of them still maintains the PO.
    \begin{proposition}
    $01 \preceq_{P,2,BMSC} 10$ and $01 \preceq_{S,2,BMSC} 10$ for $P=S=1$.
    \end{proposition}
    \pf From the upper bound and lower bound, we complete the proof by $\sqrt{2x^2-x^4}\ge x+x^2-x^3$, and $\sqrt{2x^4-x^6}\ge x^2$, $\forall x\in [0,1]$. \e

  The PO for the front insertion sequence is given below.
        \begin{lemma}\label{lemm}
        For any $p^n \in \{0,1\}^n$ and $n\ge 1$, the following equality holds for any block puncture (also for block shortening):
        \begin{equation}
        p^n 01 \preceq_{P,m,BEC} p^n 10
        \end{equation}
        \end{lemma}
        \pf Let $\bar{P}=\frac{P}{2^m}N$, $N=2^{n+2}$ and denote the binary expansion of $\bar{P}$ by $\alpha_{n+1} \cdots \alpha_1 \alpha_0$. Firstly, we can find the law: $p^n01$ and $p^n10$ are obtained from four forms after polarizing $n$ times according to $\alpha_1 \alpha_0$:
        \begin{equation}
        \begin{aligned}
        & (x,x,x,x),\alpha_1 \alpha_0=00\\
        & (x,x,x,y),\alpha_1 \alpha_0=01\\
        & (x,x,y,y),\alpha_1 \alpha_0=10\\        
        & (x,y,y,y),\alpha_1 \alpha_0=11\\
        \end{aligned}
        \end{equation}
        where $0<y<x<1$. The case $(x,x,x,x)$ is obvious from the traditional polarization and the case $(x,y,y,y)$ is similar to $(x,x,x,y)$.
        \begin{enumerate}
        \item $(x,x,x,y)$: $Z_{P,m,p^n01}=(2x-x^2)(x+y-xy)$, $Z_{P,m,p^n10}=x^2+xy-x^3y$. Then $(2x-x^2)(x+y-xy)\ge x^2+xy-x^3y$ holds for $0<y<x<1$ because $x(1-y)+y(1-x)\ge0$ for $0<y<x<1$.\\
        \item $(x,x,y,y)$: $Z_{P,m,p^n01}=(x+y-xy)^2$, $Z_{P,m,p^n10}=2xy-x^2y^2$. Then $(x+y-xy)^2\ge 2xy-x^2y^2$ holds for $0<y<x<1$ because $x^2(1-y)^2+y^2(1-x)^2\ge0$ for $0<y<x<1$.\\
        \end{enumerate}
        Hence, $p^n 01 \preceq_{P,m,BEC} p^n 10$.\e
        
        The PO for the back insertion sequence is given below.
        \begin{lemma}
        If $\alpha \preceq_{P,m,BMSC} \gamma$ holds for any block puncture, $|\alpha|=|\gamma|=n$, then $\alpha q^t\preceq_{P,m,BMSC} \gamma q^t$ holds for any block puncture and any $q^t\in \{0,1\}^t$. This is also true for shortening.
        \end{lemma}
        \pf Claim: During the $n+t$ times polarization, any pair $(x,y)$ in the symmetric position after $n$ times polarization by $\alpha$ and $\gamma$, is equal to the pair $(Z_{P,m,\alpha}(x),Z_{P,m,\gamma}(x))$ for some sequential puncture.
        
        Proof of claim: actually, we can trace the initial Bhattacharyya parameter of the $2^t$ symmetric values:
        
        For $i$-th symmetric values, $1\le i\le 2^t$: the initial Bhattacharyya parameter of the block puncture is $(z_i,z_{i+2^t},\cdots,z_{i+(2^n-1)2^t})$, where $(z_1,\cdots,z_{2^{n+t}})$ is the initial Bhattacharyya parameter of $\alpha q^t$ and $\gamma q^t$.
        
        Because $\alpha \preceq_{P,m,BEC} \gamma$ holds for any block puncture, from the claim, we know $\alpha q^t$ and $\gamma q^t$ can be regarded as the result of $t$ times polarization, where each value of the $2^t$ positions in $\alpha q^t$ is greater than the symmetric value in $\gamma q^t$. Consequently, after the back $t$ times polarization, $Z_{P,m,\alpha q^t}(x)\ge Z_{P,m,\gamma q^t}(x)$.\e
        
        \begin{corollary}\label{coro}
        For any $p^n\in \{0,1\}^n$, $q^t\in \{0,1\}^t$ and $n,t\ge 1$, the following equality holds for any block puncture (also for block shortening):
        \begin{equation}
        p^n 01 q^t \preceq_{P,m,BEC} p^n 10 q^t.
        \end{equation}
        \end{corollary}
        Then combining Lemma \ref{lemm} and Corollary \ref{coro}, we can obtain Theorem \ref{0110} by induction on $\ell$ like \cite{general}.
\begin{theorem}\label{0110}
For any $n,t,\ell \ge 0$ $p^n \in \{0,1\}^n$, $q^t \in \{0,1\}^t$, $r^\ell \in \{0,1\}^\ell$, we have
\begin{equation}
p^n0r^\ell 1q^t \preceq_{P,m,BEC} p^n1r^\ell 0q^t
\end{equation}
This is also true for $\preceq_{S,m,BEC}$.
\end{theorem}
\subsection{A Sufficient Condition for Inherited POs in BEC}
We firstly obtain a sufficient condition for inherited POs in the BEC above, and other conclusions in this subsection are obtained by verifying the sufficient condition. 

\begin{lemma}
Consider the $P/2^m$ puncturing or $S/2^m$ shortening, and given $\tau_1,\tau_2\in\{0,1\}^m$. If 
	\begin{equation}
	h_{\tau_i}:=f_{\tau_i}^{-1}\circ Z_{\tau_i},\;\;i=1,2,
	\end{equation}
	satisfy that $h_{\tau_1}(x)\le h_{\tau_2}(x), \forall x\in[0,1]$, then 
	\begin{equation}
	\tau_2\alpha\preceq_{BEC} \tau_1\gamma \Rightarrow\tau_2\alpha\preceq_{P,m,BEC} \tau_1\gamma.
	\end{equation}
	
	where $f_{\tau_i}$ is the traditional Bhattacharyya parameter defined by (\ref{bp}), and $Z_{\tau_i}$ is the rate-matched Bhattacharyya parameter defined by (\ref{punc_evo}). $\preceq_{BEC}$ is the traditional PO under the BEC \cite{newpo}.
	
\end{lemma}

\pf $Z_{\tau_2 \alpha}(x)=f_\alpha \circ f_{\tau_2}(h_{\tau_2}(x))\ge f_\alpha \circ f_{\tau_2}(h_{\tau_1}(x))\ge f_\gamma \circ f_{\tau_1}(h_{\tau_1}(x))=Z_{\tau_1 \gamma }(x).$ \e

When $P=1$, we get a inherited PO related to a form of $0^p1^q$, $p+q=m$.

\begin{theorem}
For any $1/2^m$ puncturing,
\begin{equation}
\begin{aligned}
& 0^p 1^{m-p}\alpha \preceq_{BEC} 0^q1^{m-q} \gamma \\
\Rightarrow &  0^p 1^{m-p}\alpha \preceq_{1,m,BEC} 0^q1^{m-q} \gamma,\forall 0\le q\le p\le m.\\
\end{aligned}
\end{equation}
\end{theorem}

\pf
	\begin{equation}\label{01<p01}
	\Leftarrow h_{0^{m-k-1}1^{k+1}}(x)\le h_{0^{m-k}1^k}(x),\forall x\in[0,1], m>k\in\mathbb{N}^+
	\end{equation}
	\begin{equation}
	\begin{aligned}
	& \Leftrightarrow \forall x\in[0,1],\left( 1-(1-(1-x)^{2^{m-k-1}})^{1-2^{-k-1}}\right) ^{2}\\
	& \ \ \ \ \ \ \ \ \ \ \ \ \ \ \ \ \ \ \ \ \ \ \ \ \ \ \ \ \ \ -\left( 1-(1-(1-x)^{2^{m-k}})^{1-2^{-k}}\right)\ge 0.\\
	\end{aligned}
\end{equation}

$\overset{t=(1-x)^{2^{m-k-1}}}{\Longleftrightarrow}$
\begin{equation}
\left( 1-(1-t)^{1-2^{-k-1}}\right) ^{2}-\left( 1-(1-t^2)^{1-2^{-k}}\right)\ge 0,\;\;\forall t\in[0,1].
\end{equation}
$\Leftrightarrow$
\begin{equation}
(1-t)^{2-2^{-k}} -2(1-t)^{1-2^{-k-1}}+(1-t^2)^{1-2^{-k}}\ge 0,\;\;\forall t\in[0,1].
\end{equation}
$\Leftrightarrow$
\begin{equation}
(1-t)^{1-2^{-k-1}}-2+(1+t)^{1-2^{-k}}(1-t)^{2^{-k-1}}\ge 0,\;\;\forall t\in[0,1].
\end{equation}
$\overset{u=1-t}{\Longleftrightarrow}$
\begin{equation}
u+(2-u)^{1-2^{-k}}\ge 2u^{2^{-k-1}},\;\;\forall u\in[0,1].
\end{equation}
Denote $f(u)=u+(2-u)^{1-2^{-k}}- 2u^{2^{-k-1}}$, then $f(1)=0,f'(1)=0,$
\begin{equation}
\begin{aligned}
f''(u)& =(1-2^{-k})2^{-k}(2-u)^{-2^{-k}-1}\\
& +2^{-k}(1+2^{-k-1})u^{-2^{-k-1}-2}\ge 0\\
\end{aligned}
\end{equation}
thus $f(u)\ge 0,\;\;\forall x\in[0,1]$, then \eqref{01<p01} is proved.
\e

Here is a conclusion for any block rate matching.
    
\begin{theorem}
Consider the $P/2^m$ puncturing or $S/2^m$ shortening, and $|\gamma |=m-1$, we have
\begin{equation}
\gamma 0\alpha \preceq_{BEC} \gamma 1 \beta \Rightarrow \gamma0\alpha \preceq_{P,m,BEC} \gamma 1 \beta
\end{equation}
\end{theorem}

\pf we only need to proof $h_{\gamma0}(x)\ge h_{\gamma 1}(x),\forall x\in [0,1]$.
\begin{equation}
f_{\gamma 1}^{-1}(x)=f_{\gamma}^{-1}(\sqrt{x}),f_{\gamma 0}^{-1}(x)=f_{\gamma}^{-1}(1-\sqrt{1-x}).
\end{equation}
Let

\begin{equation}
\begin{aligned}
& Z_{\gamma 1}(x)=U_\gamma (x)L_\gamma(x),\\
& Z_{\gamma 0}(x)=U_\gamma(x)+L_\gamma(x)-U_\gamma(x)L_\gamma(x).\\
\end{aligned}
\end{equation}
Then
\begin{equation}
\Leftarrow f_{\gamma}^{-1}(1-\sqrt{1-Z_{\gamma 0}(x)})\ge f_{\gamma}^{-1}(\sqrt{Z_{\gamma 1}(x)})
\end{equation}
\begin{equation}
\Leftarrow \left(Z_{\gamma 0}(x)+Z_{\gamma 1}(x) \right)^2 \ge 4 Z_{\gamma 1}(x)
\end{equation}
\begin{equation}
\Leftarrow (U_\gamma(x)-L_\gamma(x))^2\ge 0.
\end{equation}\e
	
	\section{New Partial Orders in BMSC}\label{section4}
	
In this section, we establish a general PO from BEC to BMSC under block rate matching. While the process of proof is different from that in the mother polar code. Firstly, we construct a convolution mapping in Lemma \ref{oto}. Then we use this mapping to prove that geometric mean pair exhibits a superior polarization effect in Proposition \ref{avpa}. Furthermore, we can proof a critical inequality in Lemma \ref{eq} by Proposition \ref{avpa}. Finally, utilizing the inequality and the technology of the upper and lower bounds like \cite{newpo}, we establish the general PO from BEC to BMSC under block rate matching.
	
	Before we construct the convolution mapping, we see what kind of two positions do convolution for defining the convolution mapping.
\begin{proposition}\label{convo}
Let $N=2^n$, $i,j\in \{1,2,\cdots,N\},2^s<i\le 2^{s+1},2^q<j\le 2^{q+1}$, $i< j$ then $i$ and $j$ do convolution in some layer iff 
\begin{enumerate}
\item $j-i=2^q$, if $s<q$
\item $i-2^q$ and $j-2^q$ do convolution in some layer, if $s=q$\\ 
\end{enumerate}
\end{proposition}	
Then we give the definition of convolution mapping before we construct the convolution mapping between two consecutive integer sets.
\begin{definition}
A one-to-one mapping $f:\mathcal{X}\rightarrow \mathcal{Y}$ is called a convolution mapping if for $\forall x\in \mathcal{X}$, $x$ and $f(x)$ do convolution in some layer.
\end{definition}
The convolution mapping between two consecutive integer sets is constructed as follows.
	\begin{lemma}\label{oto}
$\forall\ K\in N^+$,define $\mathcal{X}=\{1,2,\cdots,K\},\mathcal{Y}=\{K+1,K+2,\cdots,2K\}$,there exist a convolution mapping $f:\mathcal{X}\rightarrow \mathcal{Y}$.
\end{lemma}

\pf See Appendix \ref{pf of oto}. \e

An example is given to facilitate our view of the convolution mapping in the form.
\begin{example}\label{exa1}
$\mathcal{X}=\{1,2,3,4,5\}$, $\mathcal{Y}=\{6,7,8,9,10\}$, then

\begin{equation}
f(x)=\left\{\begin{aligned}
& x+2^3,when\ x\in \{1,2\}\\
& x+2^2,when\ x\in\{3,4\}\\
& x+2^0,when\ x=5\\
\end{aligned}\right.
\end{equation}
\end{example}

\begin{remark}
The parameter $k$ in the proof of Lemma \ref{oto} represents the polarization layer in the evolution of Bhattacharyya parameters. And the pair $(x,f(x))$ denotes the position corresponding to different values of Bhattacharyya parameters.
\end{remark}

Next we present a crucial polarization rule by utilizing the convolution mapping. It reveals that the more uniform the initial Bhattacharyya parameters are, the smaller the the polarized Bhattacharyya parameters are.

\begin{proposition}\label{avpa}
For initial Bhattacharyya parameters $Z=(\underbrace{a,\cdots,a}_{P} ,b,\cdots,b)$, where $a,b\in(0,1)$, $|Z|=N$, $1\le P\le N$, $Z_k$ and $Z_{k+1}$ are defined as follows. $Z_k$ have $k$ positions $\sqrt{ab}$ in $a$ positions and $k$ positions $\sqrt{ab}$ in $b$ positions of $Z$. And for $i,j\in \{1,\cdots,N\}$, if $z_i=a$ and $z_j=b$ do convolution in the outermost layer among all the $\sqrt{ab}$ pairs, then we replace $(z_i,z_j)$ by $(\hat{z}_i,\hat{z}_j)=(\sqrt{ab},\sqrt{ab})$ in $Z_k$ denoted by $Z_{k+1}$. Then we have $h(Z_{k+1})\le h(Z_{k})$.
\end{proposition}

\pf See Appendix \ref{pf of avpa}. \e

Constructing the convolution mapping to prove Proposition \ref{avpa} is to obtain the following inequality associated with Bhattacharyya parameters.

\begin{lemma}\label{eq}
For $P/2^m$ puncturing and $S/2^m$ shortening,we have
\begin{equation}
Z_{P,m,\beta}(x^2)\ge Z_{P,m,1 \beta}(x),\forall \mid \beta \mid =m
\end{equation}
\end{lemma}

\pf See Appendix \ref{pf of eq}. \e

Here is an illustrative example about Lemma \ref{eq} for understanding.

\begin{example}
Consider $1/4$ puncturing polar code, $\mid \alpha \mid=2$:
\begin{equation}
Z_{P,2,\alpha}(x^2)\ge Z_{P,2,1 \alpha}(x)
\end{equation}

where
\begin{equation}
Z_{P,2,\alpha}(x^2)=\left\{\begin{aligned}
& 1,\alpha=00\\
& 2x^2-x^4,\alpha=01\\
& x^2+x^4-x^6,\alpha=10\\
& x^6,\alpha=11\\
\end{aligned} \right.
\end{equation}

\begin{equation}
 Z_{P,2,1 \alpha}(x)=\left\{\begin{aligned}
& 1-(1-(x+x^2-x^3))^2,\alpha=00\\
& (x+x^2-x^3)^2,\alpha=01\\
& 2x^3-x^6,\alpha=10\\
& x^6,\alpha=11\\
\end{aligned} \right.
\end{equation}

\end{example}

The final step of preparation is to analyze the upper and lower bounds of Bhattacharyya parameters as discussed in \cite{newpo}.

\begin{lemma}\label{eq2}
Given BMSC $W$ with $Z(W) = x$, then for $P/2^m$ puncturing and $\mid \alpha \mid =m$,
\begin{equation}
\sqrt{Z_{P,m,\alpha} (x^2)}\le Z_{P,m}(W^\alpha)\le Z_{P,m,\alpha} (x)
\end{equation}
For $\tau=\alpha \gamma$,
\begin{equation}
\sqrt{f_\gamma \circ Z_{P,m,\alpha} (x^2)}\le Z_{P,m}(W^\tau)\le f_\gamma \circ Z_{P,m,\alpha} (x)
\end{equation}
This is also true for $S/2^m$ shortening.
\end{lemma}
\pf We proof for puncturing as an example by induction. Firstly, when $|\alpha|=1$, it is true obviously. 
\begin{enumerate}
	\item $\alpha=\gamma 0$: 
	\begin{align}
	Z_{P,m}(W^\alpha)&\overset{(a)}{\ge} \sqrt{2Z^2_{P,m}(W^\gamma)-Z^4_{P,m}(W^\gamma)}\\
	&\overset{(b)}{\ge} \sqrt{2Z_{P,m,\gamma}(x^2)-Z^2_{P,m,\gamma}(x^2)}\\
	&=\sqrt{Z_{P,m,\alpha}(x^2)}.
	\end{align}
	\begin{align}
	Z_{P,m}(W^\alpha)&\overset{(c)}{\le} 2Z_{P,m}(W^\gamma)-Z_{P,m}^2(W^\gamma)\\
	&\overset{(d)}{\le} 2Z_{P,m,\gamma}(x)-Z^2_{P,m,\gamma}(x)\\
	&=Z_{P,m,\alpha}(x).
	\end{align}
	
	\item It is obviously for the case $\alpha=\gamma 1$ because $Z_{P,m}(W^\alpha)=Z_{P,m}^2(W^\gamma)$.
	
\end{enumerate}

where $(a)$ and $(c)$ are from the lower and upper bounds \cite{general}, $(b)$ and $(d)$ are from the induction.

\e

Leveraging Lemma \ref{eq} and Lemma \ref{eq2}, we can derive our main theorem, which deriving the PO of the BMSC by leveraging the PO of the BEC under block rate matching.

\begin{theorem}\label{main1}\label{theorem1}
For block puncturing, $\mid \gamma\mid=\mid \alpha \mid\ge m$, we have 
\begin{equation}
\gamma \preceq_{P,m,BEC} \alpha \Rightarrow \gamma 1 \preceq_{P,m,BMSC} 1\alpha.
\end{equation}
This is also true for $S/2^m$ shortening.
\end{theorem}

\pf According to Lemma \ref{eq2}
\begin{equation}
\begin{aligned}
Z_{P,m,\alpha}(x^2)& \le f_1 \circ f_1^{-1}\circ Z_{P,m,\gamma} \circ f_1(x)\\
& \le f_1\circ Z_{P,m}(W^\gamma)=Z_{P,m}(W^{\gamma 1})\\
\end{aligned}
\end{equation}

And use Lemma \ref{eq}, we have

\begin{equation}
Z_{P,m}(W^{1 \alpha})\le Z_{P,m,1 \alpha}(x)\le Z_{P,m,\alpha}(x^2)\le Z_{P,m}(W^{\gamma 1})
\end{equation}

So $ \gamma 1\preceq_{P,m,BMSC} 1\alpha$.\e

The following proposition is a corollary of Theorem \ref{main1}.

\begin{proposition}
For block puncturing, $\mid \gamma\mid=\mid \alpha \mid\ge m$, we have 
\begin{equation}
1\gamma \preceq_{P,m,BEC} \alpha1 \Rightarrow \gamma \preceq_{P,m,BMSC}\alpha.
\end{equation}
This is also true for $S/2^m$ shortening.
\end{proposition}

\pf $1\gamma \preceq_{P,m,BEC} \alpha1 \Rightarrow Z_{\alpha 1}(x)\le
 Z_{1\gamma }(x),x \in [0,1] \Rightarrow Z_{\alpha}(x) \le \sqrt{Z_{1\gamma }(x)} ,x \in [0,1] \Rightarrow Z(W^\alpha)\le Z_\alpha (x)\le \sqrt{Z_{1\gamma }(x)} \le \sqrt{Z_{\gamma }(x^2)}\le Z(W^\gamma) \Rightarrow \gamma \preceq_{P,m,BMSC}\alpha$. \e
	\section{Simulation}\label{section5}
	
    When $N=1024$ and considering $1/4$ block puncturing, there are $C_{768}^2=294528$ path pairs in total. According to Theorem \ref{theorem1}, we find 198258 pairs satisfy the PO $\preceq_{1,2,BMSC}$. If the length $n_0$ of the leading identical sequence components exceed $2$, the partial order of the two sequences match that of traditional polar codes after removing the first $n_0$ bits. So in this case, we refer to \cite{newpo} to check the pair. By employing this method, we identify 212226 pairs.

    We generate the information set $\mathcal{A}_{GA}$ by GA reconstruction under rate matching \cite{b11} at $SNR=2.2dB$, and we observe that $\mathcal{A}_{GA}$ follows the partial order among all the 212226 pairs. It verifies the POs from Theorem \ref{theorem1} are beneficial for constructing block punctured polar codes. 
    
    Then we generate the information set $\mathcal{A}_{PW}$ by PW reliability sequence in \cite{beta-expansion}. And $\mathcal{A}_{improved}$ is generated by replace the positions in $\mathcal{A}_{PW}$ utilizing the PO pairs from \ref{theorem1}, which are contrary to PW sequence. Fig. 2 presents a performance comparison between the two polar codes under block puncturing and shortening. It is observed that $\mathcal{A}_{improved}$ has a gain of $0.13dB$ compared to $\mathcal{A}_{PW}$ under block puncturing. This illustrates that the PW construction can be further optimized from the perspective of PO.
    
    	\begin{figure}[!t]
		\centering
		\subfigure[]{
		\includegraphics[width=4.25cm,height=4.3cm]{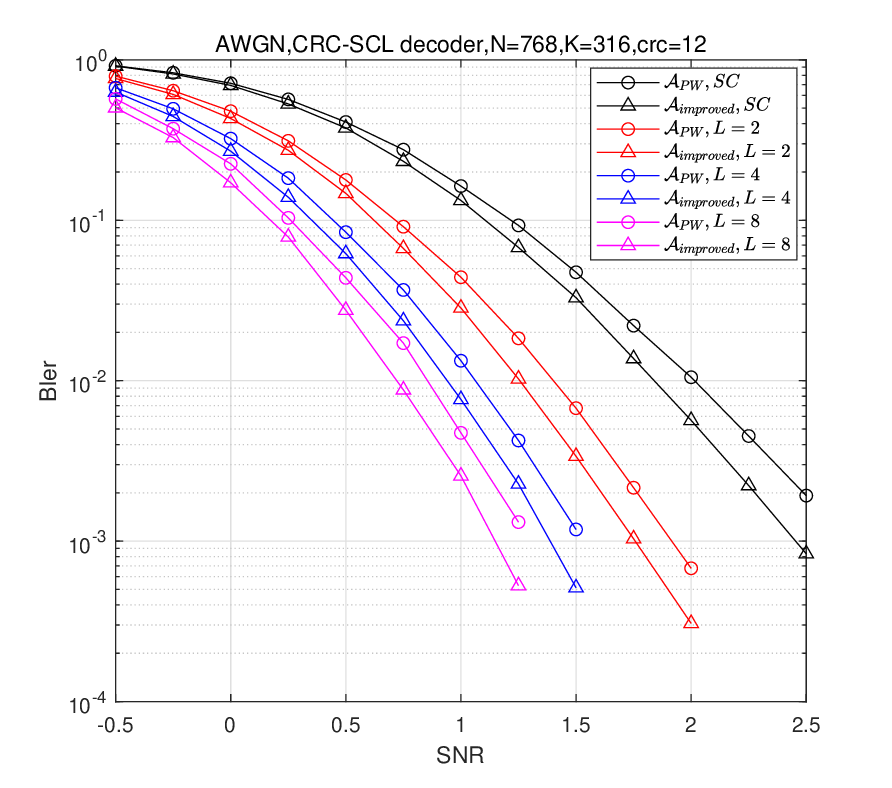}}
		\subfigure[]{
		\includegraphics[width=4.25cm,height=4.3cm]{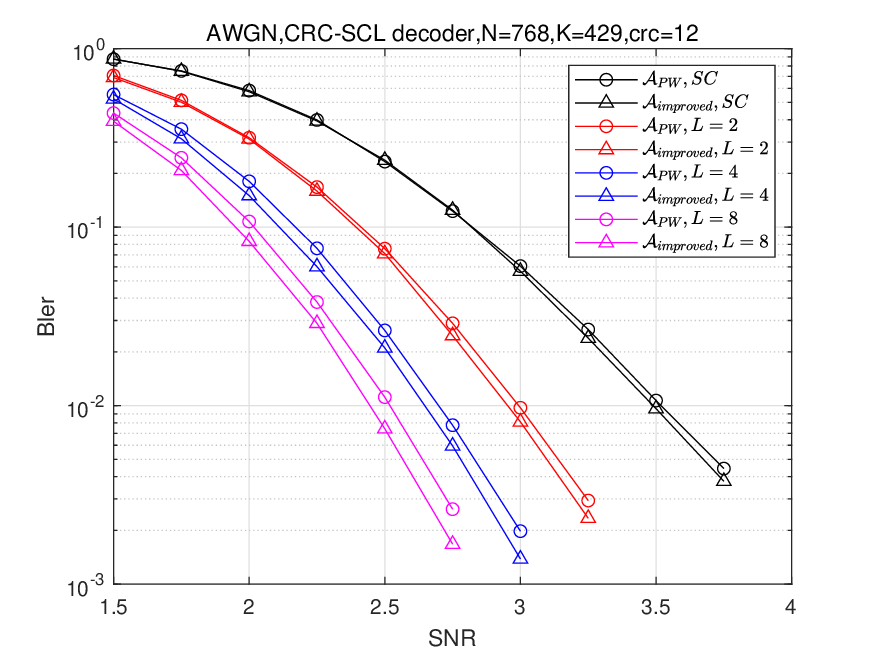}}
		\caption{SCL performance for two different information sets $\mathcal{A}_{PW}$ and $\mathcal{A}_{improved}$ with CRC length $12$ and list $L \in \{1,2,4,8\}$. }
	\end{figure}

	\section{Conclusion}\label{section6}
	In this paper, we firstly establish partial orders under block rate matching. And we introduce a sufficient condition for verifying the inherited POs of BEC under block rate matching. For the research in the BMSC, we demonstrate the property that the geometric mean of a pair of Bhattacharyya parameters decreases after polarization. By combining this result with the technique of upper and lower bounds of Bhattacharyya parameters, we establish that under block rate matching, the POs of BMSC can be derived from the POs of BEC. Finally, we verify that our work has guiding significance for the construction of polar codes under block puncturing.

\begin{appendix}
		\subsection{Proof of Lemma \ref{oto}}\label{pf of oto}
		Let $2^{q}<K\le 2^{q+1}$. In order to establish the one-to-one mapping $f$, we firstly divide $\mathcal{X},\mathcal{Y}$ into several sets $\mathcal{X}_k=\{x_k^1,\cdots,x_k^{t(k)}\},\mathcal{Y}_k=\{y_k^1,\cdots,y_k^{t(k)}\}$:

1. $\mathcal{X}_1=\{1,2\cdots,2K-2^{q+1}\}$, $\mathcal{Y}_1=\{1+2^{q+1},\cdots,2K\}$, $2^{q+1}\le y_1^{t(k)}-x_1^1=2K-1<2^{q+2}$.

2. $\mathcal{X}_k=\{x_{k-1}^{t(k-1)}+1,\cdots,y_{k-1}^{t(k-1)}-1-2^{m_k}\}$, $\mathcal{Y}_k=\{x_{k-1}^{t(k-1)}+1+2^{m_k},\cdots,y_{k-1}^{t(k-1)}-1\}$, $x_k^1\le x_k^{t(k)}<y_k^1 \Leftrightarrow 2^{m_k}\le y_{k}^{t(k)}-x_{k}^{1}<2^{m_k+1}$ until $y_k^1-x_k^{t(k)}=1\Leftrightarrow y_{k}^{t(k)}-x_{k}^{1}=2^{m_k+1}-1, $ denoted by $k_0$.

Then $\mathcal{X}=\mathop{\cup}_{k=1}^{k_0} \mathcal{X}_k,\mathcal{Y}=\mathop{\cup}_{k=1}^{k_0} \mathcal{Y}_k$,and we construct $f$ by $f(x_k^i)=y_k^i,1\le k \le k_0,1\le i\le t(k)$. $f$ is a convolution mapping because from Proposition \ref{convo}: $2^{q+1}+\mathop{\sum}_{k=1}^{k_0}(-1)^k 2^{m_k}+1$ and $2^{q+1}+\mathop{\sum}_{k=1}^{k_0+1}(-1)^k 2^{m_k}+1$ do convolution iff $1$ and $2^{m_{k_0}}+1$ do convolution; $2^{q+1}+\mathop{\sum}_{k=1}^{k_0}(-1)^k 2^{m_k}$ and $2^{q+1}+\mathop{\sum}_{k=1}^{k_0+1}(-1)^k 2^{m_k}$ do convolution iff $2^{m_{k_0}}$ and $2^{m_{k_0}+1}$ do convolution.
		\e
		\subsection{Proof of Proposition \ref{avpa}}\label{pf of avpa}
		We induct on code length $N$: assume it is right for $\frac{N}{2}$.

\noindent (i)$i\le \frac{N}{2},j>\frac{N}{2}(or\ i> \frac{N}{2},j\le\frac{N}{2})$: then $j=i+\frac{N}{2}$
\begin{equation}
 z_i\bar{*} z_j\ge 2\sqrt{z_i z_j}-z_i z_j;z_i z_j=z_i z_j
\end{equation}
it means $h(Z_{k+1}) \le h(Z_k)$.

\noindent (ii)$i\le \frac{N}{2},j\le\frac{N}{2}(or\ i> \frac{N}{2},j>\frac{N}{2})$: then $P\ne \frac{N}{2}$
\begin{equation}
\begin{aligned}
& z_i\bar{*} z_{i+\frac{N}{2}}=a+b-ab;\\
& z_j\bar{*} z_{j+\frac{N}{2}}=2b-b^2;\\
\end{aligned}
\end{equation}

\begin{equation}
\begin{aligned}
& z_i\underline{*} z_{i+\frac{N}{2}}=ab;z_j\underline{*} z_{j+\frac{N}{2}}=b^2\\
\end{aligned}
\end{equation}

\begin{equation}
\begin{aligned}
& \hat{z}_i\bar{*} z_{i+\frac{N}{2}}=\sqrt{ab}+b-\sqrt{ab}b;\\
& \hat{z}_j\bar{*} z_{j+\frac{N}{2}}=\sqrt{ab}+b-\sqrt{ab}b;\\
\end{aligned}
\end{equation}

\begin{equation}
\begin{aligned}
& \hat{z}_i\underline{*} z_{i+\frac{N}{2}}=\hat{z}_j\underline{*} z_{j+\frac{N}{2}}=\sqrt{ab}b=\sqrt{z_i z_{i+\frac{N}{2}}}\underline{*}\sqrt{ z_j z_{j+\frac{N}{2}}}\\
\end{aligned}
\end{equation}
Because for $i,j\le \frac{N}{2}$, $z_i$ and $z_j$ can not convolve in the outermost layer, none of the $\sqrt{ab}$ pairs can convolve in the outermost layer with the condition of lemma. Let $z_{up}$ and $z_{down}$ denote values in the front and back of $Z_k$ relatively, then we have the following intuition
\begin{equation}
\left\{
\begin{aligned}
& z_{up}\in \{a,b,\sqrt{ab}\},z_{down}=b,when\ P< \frac{N}{2},i\le \frac{N}{2},j\le\frac{N}{2}\\
& z_{up}\in \{a\},z_{down}\in \{a,b,\sqrt{ab}\},when\ P> \frac{N}{2},i> \frac{N}{2},j>\frac{N}{2}\\
\end{aligned}\right.
\end{equation}
On the one hand,
\begin{equation}
\left\{
\begin{aligned}
& Z^{(1)}_{back}=(ab,\cdots,ab,b^2,\cdots,b^2),when\ P< \frac{N}{2},i\le \frac{N}{2},j\le\frac{N}{2}\\
& Z^{(1)}_{back}=(a^2,\cdots,a^2,ab,\cdots,ab),when\ P> \frac{N}{2},i> \frac{N}{2},j>\frac{N}{2}\\
\end{aligned}\right.
\end{equation}
There are $k$ positions containing $\sqrt{ab}b\ (or\ \sqrt{ab}a)$ instead of $ab$ and $k$ positions containing $\sqrt{ab}b\ (or\ \sqrt{ab}a)$ instead of $b^2\ (or\ a^2)$ in $Z^{(1)}_{back}$.

And $\hat{Z}^{(1)}_{back}$ is generated by replacing a pair of $Z^{(1)}_{back}$ as follows:
\begin{equation}
\left\{
\begin{aligned}
\hat{Z}^{(1)}_{back}:& replace\ (z^{(1)}_i,z^{(1)}_j)\ by\ (\sqrt{ab}b,\sqrt{ab}b),\\
& when\ P< \frac{N}{2},i\le \frac{N}{2},j\le\frac{N}{2}\\
\hat{Z}^{(1)}_{back}:& replace\ (z^{(1)}_{i-\frac{N}{2}},z^{(1)}_{j-\frac{N}{2}})\ by\ (\sqrt{ab}a,\sqrt{ab}a),\\
& when\ P> \frac{N}{2},i> \frac{N}{2},j>\frac{N}{2}\\
\end{aligned}\right.
\end{equation}
We know that all the averaged pairs either belong to the front half or the back half, so $(z^{(1)}_i,z^{(1)}_j)$ or $(z^{(1)}_{i-\frac{N}{2}},z^{(1)}_{j-\frac{N}{2}})$ remains in the outermost layer among all the averaged pairs. By induction,we have
\begin{equation}
h(\hat{Z}^{(1)}_{back})\le h(Z^{(1)}_{back})
\end{equation}
On the other hand,
\begin{equation}
\left\{
\begin{aligned}
Z^{(1)}_{front}=& (a+b-ab,\cdots,a+b-ab,2b-b^2,\cdots,2b-b^2),\\
& when\ P< \frac{N}{2},i\le \frac{N}{2},j\le\frac{N}{2}\\
Z^{(1)}_{front}=& (2a-a^2,\cdots,2a-a^2,a+b-ab,\cdots,a+b-ab),\\
& when\ P> \frac{N}{2},i> \frac{N}{2},j>\frac{N}{2}\\
\end{aligned}\right.
\end{equation}

There are $k$ positions containing $\sqrt{ab}+b-\sqrt{ab}b$ (or $\sqrt{ab}+a-\sqrt{ab}a$) instead of $a+b-ab$ and $k$ positions containing $\sqrt{ab}+b-\sqrt{ab}b$ (or $\sqrt{ab}+a-\sqrt{ab}a$) instead of $2b-b^2\ (or\ 2a-a^2)$ in $Z^{(1)}_{front}$.

Similarly, $\hat{Z}^{(1)}_{front}$ is generated by replacing a pair of $Z^{(1)}_{front}$ as follows:

\begin{equation}
\left\{
\begin{aligned}
\hat{Z}^{(1)}_{front}:& when\ P< \frac{N}{2},i\le \frac{N}{2},j\le\frac{N}{2},replace\ (z^{(1)}_i,z^{(1)}_j)\\
& by\ (\sqrt{ab}+b-\sqrt{ab}b,\sqrt{ab}+b-\sqrt{ab}b)\\
\hat{Z}^{(1)}_{front}:& when\ P> \frac{N}{2},i> \frac{N}{2},j>\frac{N}{2},replace\ (z^{(1)}_{i-\frac{N}{2}},z^{(1)}_{j-\frac{N}{2}})\\
& by\ (\sqrt{ab}+a-\sqrt{ab}a,\sqrt{ab}+a-\sqrt{ab}a)\\
\end{aligned}\right.
\end{equation}
As a medium step, we generate $\bar{Z}_{front}^{(1)}$ by replacing the same pair of $Z^{(1)}_{front}$ as follows:
\begin{equation}
\left\{
\begin{aligned}
\bar{Z}^{(1)}_{front}:& when\ P< \frac{N}{2},i\le \frac{N}{2},j\le\frac{N}{2},replace\ (z^{(1)}_i,z^{(1)}_j)\\
& by\ (\sqrt{(a+b-ab)(2b-b^2)},\sqrt{(a+b-ab)(2b-b^2)})\\
\bar{Z}^{(1)}_{front}:& when\ P> \frac{N}{2},i> \frac{N}{2},j>\frac{N}{2},replace\ (z^{(1)}_{i-\frac{N}{2}},z^{(1)}_{j-\frac{N}{2}})\\
& by\ (\sqrt{(a+b-ab)(2a-a^2)},\sqrt{(a+b-ab)(2a-a^2)})\\
\end{aligned}\right.
\end{equation}

According to $(z^{(1)}_i,z^{(1)}_j)$ or $(z^{(1)}_{i-\frac{N}{2}},z^{(1)}_{j-\frac{N}{2}})$ remains in the outermost layer among all the averaged pairs, and
\begin{equation}
\left\{
\begin{aligned}
& \sqrt{(z_i\bar{*} z_{i+\frac{N}{2}})(z_j\bar{*} z_{j+\frac{N}{2}})}=\sqrt{(a+b-ab)(2b-b^2)},P<\frac{N}{2}\\
& \sqrt{(z_i\bar{*} z_{i-\frac{N}{2}})(z_j\bar{*} z_{j-\frac{N}{2}})}=\sqrt{(a+b-ab)(2a-a^2)},P>\frac{N}{2}\\
\end{aligned}
\right.
\end{equation}
Then by the induction, we conclude that 
\begin{equation}
h(\bar{Z}^{(1)}_{front})\le h(Z^{(1)}_{front})
\end{equation}
And use the inequality
\begin{equation}
\sqrt{ab}+b-\sqrt{ab}b\le \sqrt{(a+b-ab)(2b-b^2)},\forall a,b\in[0,1]
\end{equation}
then we have
\begin{equation}
h(\hat{Z}^{(1)}_{front})\le h(\bar{Z}^{(1)}_{front})
\end{equation}
So
\begin{equation}
h(\hat{Z}^{(1)}_{front})\le h(\bar{Z}^{(1)}_{front})\le h(Z^{(1)}_{front})
\end{equation}
Finally,
\begin{equation}
\begin{aligned}
h(Z_{k+1})& =(h(\hat{Z}^{(1)}_{front}),h(\hat{Z}^{(1)}_{back})) \\
& \le (h(Z^{(1)}_{front}),h(Z^{(1)}_{back}))= h(Z_k)\\
\end{aligned}
\end{equation}\e

		\subsection{Proof of Lemma \ref{eq}}\label{pf of eq}
For simplification, denote the initial Bhattacharyya parameters of $Z_{P,m,\beta}(x^2)$ by $Z$, and one time down polarization of the initial Bhattacharyya parameters of $Z_{P,m,1 \beta}(x)$ by $\hat{Z}$. Then $Z$ and $\hat{Z}$ can be written as
\begin{equation}
\left\{\begin{aligned}
P\le \frac{N}{2}:& Z=(\underbrace{1,\cdots,1}_{P} ,\underbrace{x^2,\cdots,x^2}_{P},\underbrace{x^2,\cdots,x^2}_{N-2P}),\\
& \hat{Z}=(\underbrace{x,\cdots,x}_{2P} ,\underbrace{x^2,\cdots,x^2}_{N-2P})\\
P> \frac{N}{2}:& Z=(\underbrace{1,\cdots,1}_{2P-N},\underbrace{1,\cdots,1}_{N-P},\underbrace{x^2,\cdots,x^2}_{N-P}),\\
& \hat{Z}=(\underbrace{1,\cdots,1}_{2P-N},\underbrace{x,\cdots,x}_{2(N-P)})\\
\end{aligned}\right.
\end{equation}
According to Lemma \ref{oto}, when $P\le \frac{N}{2}$, the first $2P$ numbers of $Z$ can be partitioned into $P$ pairs. When $P>\frac{N}{2}$, the symmetry between the front and back positions ensures the validity of this partitioning. It is essential to highlight that these $P$ pairs need to be arranged in ascending order of layers to satisfy the condition of Proposition \ref{avpa}.

Then we replace each $(1,x^2)$ pair in $Z$ with $(x,x)$ in turn among the $P$ pairs. Let $Z_k$ represent the initial Bhattacharyya parameters with $k$ averaged pairs, where the first $k-1$ averaged pairs are identical to those of $Z_{k-1}$.

Use Proposition \ref{avpa} we get
\begin{equation}
h(Z_{k-1}) \le h(Z_{k}),\forall 1\le k\le P
\end{equation}
So
\begin{equation}
h(\hat{Z}):=h(Z_P)\le h(Z_{P-1}) \le \cdots h(Z_{1})\le h(Z_0)=:h(Z)
\end{equation}
It means
\begin{equation}
Z_{P,m,\beta}(x^2)\ge Z_{P,m,1 \beta}(x),\forall \mid \beta \mid =m
\end{equation} \e
\end{appendix}
	
\end{document}